\documentstyle{mn}
\input{epsf}

\title[Properties of observed Ly-$\alpha$ forest]
{Properties of observed Ly-$\alpha$ forest}

\author[Demia\'nski \& Doroshkevich]
       {M. Demia\'nski$^{1,2}$,  A.G. Doroshkevich$^{3,4}$,
	\& V.Turchaninov$^4$\\
	$1$Institute of Theoretical Physics,
                       University of Warsaw,
                       00-681 Warsaw, Poland\\
        $2$Department of Astronomy, Williams College,
           Williamstown, MA 01267, USA\\
	$3$Theoretical Astrophysics Center,
          Juliane Maries Vej 30,
          DK-2100 Copenhagen \O, Denmark\\
	$4$Keldysh Institute of Applied Mathematics,
                        Russian Academy of Sciences,
                        125047 Moscow,  Russia\\
}

\date{Accepted ...,
      Received ...,
	in original form ... .}

\begin{document}
\maketitle

\begin{abstract}
The main observed properties of Ly-$\alpha$ absorbers are
investigated on the basis of theoretical model of formation 
and evolution of DM structure elements. This model is generally 
consistent with simulations of absorbers formation and with 
statistical description of structure evolution based on the 
Zel'dovich theory. The analysis of redshift variations of 
comoving linear number density of absorbers was performed 
in our previous paper.

We show that the observed characteristics of Doppler parameter
can be related to the size of DM structure elements what
allows us to explain the observed distribution of Doppler
parameter. This distribution is found to be consistent with
the Gaussian initial perturbations. The observed
characteristics of entropy and column density, $N_{HI}$,
confirm that  merging of pancakes is the main evolutionary
process at redshifts $z\geq 2$. The observed sample of absorbers
characterizes mainly the matter distribution within large
low density regions and therefore it is difficult to
reconstruct the density field from the distribution of
absorbers.
\end{abstract}

\begin{keywords}  cosmology: large-scale structure of the Universe ---
          quasars: absorption: general --- surveys.
\end{keywords}

\section{Introduction}

The intergalactic nature of Ly-$\alpha$ forest was established
by Sargent et al. (1980) and later many models of absorbers
formation and evolution were proposed (see, e.g., Rees 1986,
1995; Ikeuchi \& Ostriker 1986; Bond, Szalay \& Silk 1988;
Miralda-Escude et al. 1996; Hui et al. 1997; Nath 1997; Valageas, 
Schaeffer, \& Silk 1999). The essential progress was reached 
through simulations of dynamical and thermal evolution of gaseous 
component with the CDM-like power spectrum which is probably 
responsible for the formation of observed galaxy distribution. 
Such simulations (Petitjean et al. 1995; Hernquist et al. 1996; 
Bond \& Wadsley 1997; Zhang et al. 1997, 1998; Theuns et al. 1998, 
1999; Bryan et al. 1999; Dav\'e et al. 1999; Weinberg et al. 1998; 
Machacek et al. 2000) reproduce successfully the main observed 
properties of absorbers and connect this problem with a more 
general problem, that is the nonlinear evolution of initial 
perturbations and formation of large scale matter and observed 
galaxy distribution. This progress allows us to consider the 
properties of absorbers in the context of nonlinear theory of 
gravitational instability and statistical description of formation 
and evolution of DM structure (Zel'dovich 1970; Shandarin \& 
Zel'dovich 1988; Demia\'nski \& Doroshkevich 1999, hereafter 
DD99; Demia\'nski et al. 2000, hereafter DDMT).

It is commonly recognized that in the DM dominated universe the
fundamental properties of observed matter distribution are determined 
by the evolution of DM structure elements. The approximate statistical
description of expected characteristics of DM structure elements
based on the Zel'dovich theory was given in DD99 and DDMT for 
the CDM-like transfer function (Bardeen et al. 1986) and the 
Harrison -- Zel'dovich initial power spectrum. This process can 
be outlined as a random formation and merging of Zel'dovich 
pancakes, their successful transformation to filamentary component 
of structure, and the hierarchical merging of both pancakes and 
filaments to form rich walls. All steps of this evolution are driven 
by the initial power spectrum. 

Some results of this statistical description were compared with 
simulations at small redshifts (DD99, Doroshkevich et al., 1999, 
hereafter DMRT; DDMT). This comparison shows that the main 
simulated and observed structure characteristics are consistent 
with the theoretical expectations. Here we use this approach for 
the analysis and interpretation of the absorbers observed at 
large redshifts as a Ly$-\alpha$ forest, and we show that 
some of the observed characteristics of Ly-$\alpha$ absorbers can 
also be successfully described in the framework of this 
theoretical model. 

The validity of this description, when it is applied to structure 
at high redshifts, is not yet reliably verified with available 
simulations due to a small density contrast of poor structure 
elements dominating at higher redshifts. The first tests revealed, 
however, the existence of three 
kinds of structure elements, namely, high density filaments, 
clumps, and low density pancakes in DM spatial distribution at 
$z=$ 3. The theoretical description (DD99) confirms also the 
self-similar character of structure evolution (at least when 
the Zel'dovich approximation can be applied). More detailed 
statistical comparison of absorbers characteristics simulated at 
high redshifts with theoretical expectations is however required.  

Such approach implies more traditional investigation of properties
of discrete absorbers rather then a continuous non-linear line-of-sight 
density field (see, e.g., Weinberg et al. 1998) and, so, we 
will concentrate more attention on statistics of discrete absorbers. 
The observations of galaxies at large redshifts verifies that  
strong nonlinear compression of DM and gaseous components occurs 
even at $z\approx$ 3 -- 5 and earlier (Steidel et al. 1998; Dey \& 
Chaffee 1998; Fan et al. 2000) and, so, 
discrete high density absorbers can already exist 
at such redshifts. This approach allows us to reach more clarity 
in the description of formation and evolution of structure and
to establish correlations between properties of observed
absorbers and invisible DM component.

It can be expected that absorbers are predominantly related to the more
numerous population of moderately rich pancakes and to periphery of
richer pancakes and filaments. This means that information obtained
from such traditional methods of analysis is related with the
extended lower density regions of the universe rather then with the
rich wall-like condensations. In this respect, the absorbers
characteristics are complementary to the information obtained from the
analysis of large scale galaxy distribution. In order to discriminate
between properties of absorbers that can be associated with the
evolution of DM component and are specific of the gaseous component
it is necessary to compare the expected characteristics of DM
structure and observed discrete Ly-$\alpha$ absorbers. Thus, some
fraction of  weaker absorbers formed within expanded regions is not
connected  with the DM structure (Bi \& Davidsen 1997; Zhang et al.
1998; Dav\'e  et al. 1999). Such comparison reveals also the potential
and  limitations of this approach. 

This approach relies on identification of separate absorbers in the 
observed spectra what restricts the number of available spectra.
Moreover, properties of some of the identified lines can be distorted 
due to superposition of several lines and cannot be reliably 
discriminated from the influence of diffuse intergalactic gas 
(McGill 1990; Levshakov \& Kegel 1996, 1997) what introduces some 
additional 
uncertainties in our analysis. Our study of simulations at small 
redshifts (DDMT) shows, however, that the influence of the last 
effect depends on the power spectrum of primordial perturbations 
and for the CDM-like power spectrum, at moderate redshifts, its 
impact is not very strong. This problem should be investigated in 
more details using the available simulations of absorbers.

The redshift dependence of linear number density of absorbers was
discussed in Demia\'nski, Doroshkevich \& Turchaninov (2000,
hereafter Paper I) under the assumption that the neutral hydrogen
traces the potential wells formed by DM pancakes. This model is
similar, in some respects, to previously discussed theoretical and
simulated models referred above.  In this paper we show that this
model provides a reasonable  self consistent description and
interpretation of other observed properties of absorbers and
demonstrates that merging of earlier formed structure elements plays
an important role in the evolution  of  absorbers. This approach makes
it possible to approximately discriminate   between the evolution of
DM structure elements and faster and randomly perturbed evolution of
gaseous component, that is suitably described by the evolution of
entropy of compressed gas, and to specify the main factors
responsible for it. In particular, we can roughly discriminate
between adiabatic and shock formation of absorbers and show that the
adiabatic processes may be not very important,  at least for the
formation of  stronger {\it observed} absorbers. Properties of
observed absorbers demonstrate the important role  of merging of
earlier formed structure elements for the absorber  evolution. Our
method gives reasonable fits for the observed distribution of
Doppler parameter, $b$, and hydrogen column density, $N_{HI}$, and
connects them with the basic cosmological parameters,
$\Omega_m~\&~h$, and the amplitude of initial perturbations. Our
main results are  consistent with conclusions of Zhang et al.
(1998), Weinberg et  al. (1998) and Dav\'e et al. (1999). Some
differences between the theoretical expectations and simulations
will be discussed below. 

The theory cannot yet describe in details the relaxation of
compressed matter, the disruption of structure elements caused by
the gravitational instability of compressed DM and the distribution
of neutral hydrogen across DM pancakes. Therefore, in this paper
several parameters characterizing the properties of
DM and neutral hydrogen distribution remain undetermined. They
can be estimated by applying the discussed methods to simulations 
that provide an unified physical picture of absorber formation 
and evolution.

The composition of observed absorbers is complicated and if at low
redshifts a significant number of stronger Ly-$\alpha$ lines and metal
systems is associated with galaxies (Bergeron et al. 1992; Lanzetta et
al. 1995; Cowie et al. 1995; Tytler et al. 1995; Le Brune et al. 1996)
then the population of weaker absorbers dominates at higher redshifts
and mainly disappears at redshift $z\leq$ 2. It is not observed by
other methods and can be associated with the population of weaker
structure elements formed by non luminous baryonic and DM components
and situated in extended lower density regions. Some
number of weak Ly-$\alpha$ lines observed even at small redshifts
far from galaxies (Morris et al. 1993; Stocke et al. 1995; Shull et
al. 1996) can be considered as a trace of this population.

An interesting problem arises, namely the possible 
reconstruction of spatial DM distribution using the redshift
distribution of absorbers. Two methods of such reconstruction were
proposed by Weinberg et al. (1998) and Nusser \& Haehnelt (1998). 
Here we examine the method based on one dimensional smoothing of
density field. We show that results depend strongly on the used sample 
of absorbers and on the method of identification of properties of DM 
component of absorbers. Interpretation of results of such 
reconstruction is now  questioned and more detailed investigation 
of this problem is required.

Simulations of structure evolution take into account the impact of
many important factors together and provide an unified picture of
absorber formation and evolution. But so far such simulations can 
be performed only in small boxes what introduces artificial cutoffs
in the power spectrum and makes the investigation of large scale 
structure evolution difficult. Such simulations cannot yet reproduce
all important features of interactions of small and large scale
perturbations and the direct analysis of the observed absorbers
characteristics might now be more perspective in this respect. Indeed,
the large scale modulation of redshift distribution of Ly-$\alpha$
lines found by Cristiani et al. (1996) and strong nonhomogeneities
found at $z\leq$ 2 by Williger et al. (1996), Quashnock et al.
(1996, 1998), and Connolly et al. (1996) could be attributed to the
extremely rich structure elements which are not yet found in
simulations.

This paper is organized as follows. The theoretical model of
the structure evolution is discussed in Secs. 2. Sec. 3 contains
information about the used observational databases. The results
of statistical analysis are given in Sec. 4, 5~\&~ 6. Discussion
and conclusion can be found in Sec. 7.

\section{Model of structure evolution.}

The main observational characteristics of absorption lines
are the redshift, $z_{abs}$, the column density of neutral
hydrogen, $N_{HI}$, and the Doppler parameter, $b$. On the
other hand, the theoretical description of structure
formation and evolution is dealing with the linear number
density, $n_{abs}(z)$, temperature, $T$, density of DM and
entropy of gaseous components, and with the ionization
degree of hydrogen. To connect these theoretical and observed
parameters a physical model of absorbers formation and evolution
is required. 

Broad set of such models was discussed during last twenty years 
(see references above). Here we repeat some of the assumptions 
already discussed in earlier publications. Our consideration 
is based on the statistical description of formation and evolution 
of DM structure in CDM-like models (DD99, DDMT), and it is compared 
with observed and simulated spatial distributions of DM component 
and galaxies at small redshifts.   

\subsection{Physical model of absorbers.}

Here we consider a simple self-consistent model of the absorbers
formation and evolution based on the Zel'dovich approximation. We
assume that:
\begin{enumerate}
\item{} The DM distribution forms an interconnected structure
of sheets (Zel'dovich pancakes) and filaments, their main parameters
are approximately described by the Zel'dovich approximate theory
applied to CDM-like initial power spectrum (DD99, DDMT). The richer 
DM pancakes can be relaxed, long-lived, and approximately stationary.
\item{} Gas is trapped in the gravitational potential wells formed 
by the DM distribution. The gas temperature and observed Doppler 
parameter, $b$, trace the depth of the DM potential wells. 
\item{} For a given temperature the gas density within the
wells is determined by the gas entropy created during the previous
evolution. The gas entropy changes, mainly, due to the shocks
heating in the course of merging of pancakes and, possibly,
due to the bulk heating produced by local sources.
\item{} The ionization of the gas is caused by the outer radiation
field and for the majority of absorbers ionization equilibrium
is assumed.
\item{} The evolution of observed properties of absorbers is
mainly caused by merging, transversal compression and/or expansion
and disruption of DM pancakes. Possible bulk heating of the
trapped gas and possible variations of the intensity and spectrum
of the ionizing UV radiation field can be also important and will
provide the fine-tuning between the observed and expected properties
of absorbers.
\item{} In the context of the simple model we identify the velocity 
dispersion of DM component compressed within pancakes with the 
temperature of hydrogen and the Doppler parameter $b$ of absorbers. 
We consider the possible macroscopic motion within pancakes as 
subsonic and assume that they cannot essentially distort the 
measured Doppler parameter. 
\end{enumerate}

The formation of sheet-like DM structure elements (Zel'dovich 
pancakes) as an inevitable first step of evolution of small DM 
perturbations was certainly established both by theoretical 
considerations (Zel'dovich 1970; Shandarin \& Zel'dovich 1989; 
in DD99 for CDM-like power spectrum) and numerically (Shandarin 
et al. 1995). Here we will restrict our consideration to the 
subpopulation of slowly evolving DM pancakes when their column 
density remains almost the same during the time comparable with 
$H^{-1}(z)$. In the opposite case, when rapid expansion of matter 
in the transversal directions takes place, the pancake is eroded 
and $N_{HI}$ decreases below the observational limit. The rapid 
compression transforms pancakes into filaments that is another 
subpopulation of observed Ly-$\alpha$ absorbers. These short-lived 
pancakes can be mainly identified with a subpopulation of weaker 
absorbers with a column density $N_{HI}\leq 10^{13}cm^{-2}$ 
dominated at higher redshifts $z\geq$3. 

The subpopulation of weaker absorbers also  contains 
"artificial" caustics (McGill 1990) and absorbers identified 
with slowly expanding underdense regions (Bi \& Davidsen 1997; 
Zhang et al. 1998; Dav\'e et al. 1999). These kinds of absorbers 
are not connected with DM structure and produce noise, which is 
stronger at higher redshifts $z\geq$ 3. 

Further on, even our approximate consideration cannot be applied 
to filaments and high density clumps, when both the gravitational 
potential and the gas temperature along a line of sight depend 
essentially on the matter distribution across this line. So, our 
investigation has to be restricted to the subpopulation of DM 
composed sheet-like structure elements. This means that the 
appropriate subsample of observed absorbers has to be 
considered.

Fortunately, the correlations between observed parameters $b$ 
and $N_{HI}$ and evolutionary rate of observed linear number 
density of absorbers discussed in Paper I allows us to discriminate 
statistically the filamentary and sheet-like dominated subpopulations 
of absorbers and to perform approximately such selection. Thus, 
it may be expected that the sheet-like DM composed absorbers dominate 
for $b\geq$ 17 -- 20km/s and $10^{13}cm^{-2}\leq N_{HI}\leq 
10^{14}cm^{-2}$. 

Both theoretical analysis and simulations show the successive 
transformation of sheet-like elements into filamentary-like ones 
and, at the same time, the merging of both sheet-like and 
filamentary elements into richer sheets or walls. Such continual 
transformation of structure goes on all the time. These processes 
imply the existence of complicated time-dependent internal 
structure of high density elements and, in particular, the 
essential arbitrariness in discrimination of such elements 
into filaments and sheets. The morphology of structure elements 
can be quantitatively characterized with new powerful techniques 
such as the Minimal Spanning Tree analysis (DMRT, DDMT) and 
the Minkowski Functional (Sathyaprakash et al. 1998; Kerscher 
1999). Such an analysis applied to observed and simulated 
catalogues demonstrates the continual distribution of morphological 
characteristics with a relatively small fraction of distinct 
high density filaments and elliptical clumps, and allows to 
estimate the degree of filamentarity and sheetness of both 
individual elements and sample as a whole.  

In this paper, as was noted above, we use the term 'pancake' for
structure elements with relatively small gradient of properties
(first of all temperature) across a line of sight. With such a
criterion, the anisotropic halo of filaments and clumps can also 
be considered as 'pancake-like'. At the same time, as was discussed 
in Paper I, the evolutionary  rate of observed linear number 
density of such absorbers can be similar to that typical for 
filaments or clumps. By imposing some restrictions on the observed 
$b$ and $N_{HI}$ it is possible to improve the statistical 
discrimination of these components, but even then the selection 
is not unique. More detailed investigation of observed and 
simulated properties of absorbers is required to improve the 
selection criteria and to prepare more adequate physical model 
of absorbers.   

\subsection{DM structure elements and Doppler parameter.}

Among the observed characteristics of Ly-$\alpha$ absorbers the
Doppler parameter, $b$, is more closely linked with properties
of DM component. In this section we introduce some relations
between characteristics of DM pancakes and the $b$ parameter based
both on the theoretical arguments and analysis of simulated DM
and observed  galaxy distribution at small redshifts. Some
properties of such high density DM walls formed at the redshift
$z\ll$1 were analyzed in DMRT, DD99 and DDMT. In this Section 
relations between basic characteristics of DM pancakes are 
introduced (without proofs) as a basis for further analysis. 

The fundamental characteristics of DM pancakes are the dimensionless
Lagrangian thickness, $q$, and the DM column density, $\mu_f$:
$$\mu_f\approx {\rho_f l_0q\over (1+z_f)} = {3H_0^2\over8\pi G}l_0
\Omega_m(1+z_f)^2q, \eqno(2.1)$$
$$l_0\approx {6.6\over h\Omega_m}h^{-1}{\rm Mpc}={59.4
{\rm Mpc}\over \Theta_m},~~\Theta_m = 9\Omega_mh^2,$$
where $\Omega_m$ is the dimensionless mean matter density of the
universe and $H_0$=100 h km/s/Mpc is the Hubble constant. The
Lagrangian thickness of a pancake, $l_0q$, is defined as an 
unperturbed distance at redshift $z=0$ between positions of
DM particles bounding the pancake.

\subsubsection{DM pancakes in Zel'dovich approximation.}

The expected probability distribution function (PDF) for the
Lagrangian thickness of a pancake, $q$, can be written (DD99) as
$$N_q\approx {1\over 4\tau^2\sqrt{\pi}}\cdot e^{-\xi}~{{\rm erf}
(\sqrt{\xi})\over\sqrt{\xi}} ,\quad \xi={q\over 8\tau^2},\eqno(2.2)$$
$$<\xi>=0.5+1/\pi\approx 0.82,\quad \langle \xi^2\rangle =
0.75+2/\pi\approx 1.39$$
where the dimensionless 'time' $\tau(z)$ describes the evolution
of perturbations in the Zel'dovich theory (Appendix A). More
details are given in DMRT, DD99 and DDMT.

In the Zel'dovich theory the Lagrangian thickness of DM pancake, 
$q$, is closely linked to the velocity of infalling matter, 
$v_{inf}$, the thickness, $h_{DM}$, and overdensity, $\delta_{DM}$, 
of compressed dark matter (DDMT):
$$\langle v_{inf}\rangle\approx {H(z)l_0\beta(z)\over 2(1+z)}q,
 ~~\sigma_{inf}\approx {l_0 H(z)\over (1+z)}
\tau[1+\beta(z)]\sqrt{q\over 3},$$
$$\beta(z) = {1+z\over\tau}\left|{d\tau\over dz}\right|,\eqno(2.3)$$
$$\langle h_{DM}\rangle\ll\sqrt{\langle h_{DM}^2\rangle}\approx
{2l_0\tau\over 1+z }\sqrt{q},\quad
\delta_{DM}\approx {\sqrt{q}\over 2\tau}.$$
The PDF of infalling velocity is Gaussian for Gaussian initial
perturbations (DD99, DDMT) with the mean value and dispersion
as given by (2.3).

Further on we will identify the kinetic energy accumulated by
the DM pancake with the Doppler parameter, $b_{DM}$, and with 
the {\it observed} Doppler parameter, $b$. We will assume that
$$b_{DM}^2 = \langle v_{inf}^2\rangle\approx {l_0^2 H^2(z)q\over
12 (1+z)^2}(q\beta^2(z)+4\tau^2[1+\beta(z)]^2).    \eqno(2.4)$$
This assumption is valid during some time after formation of the
pancake and allows us to reasonably describe the observed properties 
of absorbers. Later on, other processes such as the relaxation 
and small scale clustering of compressed matter, as well as, the 
pancake compression and/or expansion in transversal directions 
become important. Some of them will be discussed below.

For larger redshifts, $z\geq$ 2, $q\ll$ 1, we can introduce more
suitable notation separating out large numerical factors. We
can take with a reasonable precision
$$H(z)\approx H_0(1+z)^{3/2}\sqrt{\Omega_m}\approx 267\zeta^{3/2}
\sqrt{\Theta_m} km/s/{\rm Mpc},$$
$$\tau\approx 0.06\zeta^{-1}\tau_z,~~ \beta\approx 1,
~~\zeta=0.25(1+z),					\eqno(2.5)$$
what allows us to rewrite (2.3) and (2.4) more transparently as
$$b_{DM}\approx b_0\sqrt{\xi^2+2\xi},\quad \xi=\eta^2(1+\sqrt{1+
\eta^2})^{-1},\quad \eta=b_{DM}/b_0, $$
$$\langle b_{DM}\rangle\approx 1.43 b_0,\quad
\langle b^2_{DM}\rangle\approx 3b_0^2,\quad \delta_{DM}\approx
\sqrt{2\xi},						\eqno(2.6)$$
$$b_0 = 33km/s \zeta^{-3/2}\tau_z^2\Theta_m^{-1/2}\Theta_v,$$
where $\xi$ was introduced by (2.2), and the factor
$\Theta_v\sim$1 describes differences between (2.4) and (2.6).

These relations connect the velocity dispersion within DM
pancakes with their DM column densities and allow to obtain
(using the PDF (2.2)~) the expected PDF for $b_{DM}$ as follows:
$$N_b = N_q\cdot {dq\over db_{DM}} = {2\over b_0\sqrt{\pi}}
~e^{-\xi}{\sqrt{2+\xi}\over 1+\xi}~{\rm erf}(\sqrt{\xi}).\eqno(2.7)$$

\subsubsection{Relaxation of DM pancakes.}

The analysis of simulations (DDMT) shows that in rich pancakes
the DM particles are relaxed and gravitationally confined, and
such pancakes are long-lived and (quasi)stationary. The relaxation
of DM particles is essentially accelerated by the small scale
clustering of compressed matter and leads to evaporation of
particles with larger velocities, what restricts the observed
Doppler parameter of rich pancakes. Thus, for observed and 
simulated galaxy walls, at small redshifts, $b\sim$ 300 -- 
350km/s is found to be more typical. 

The approximate relation for the velocity dispersion of DM 
particles within relaxed pancakes can be written as follows:
$$b_{DM}\approx \epsilon(z)\sqrt{\langle b^2_{DM}\rangle}
\left({q\over\langle q\rangle}\right)^\gamma\approx
\epsilon(z)\sqrt{3}\left({q\over\langle q\rangle}\right)^\gamma,
						\eqno(2.8)$$
$$\langle \epsilon\rangle\approx
0.3 - 0.7,\quad \gamma\approx 0.6 - 0.7 .$$
Here the random dimensionless parameter $\epsilon(z)$ describes
the lost of energy in the course of relaxation. Estimates of the   
factors $\gamma$ and $\epsilon$ can be refined through a more
detailed comparison with simulations.

For relaxed pancakes, instead of (2.6) \& (2.7) we obtain
that, for example, for $\gamma=2/3$, the expected
characteristics of Doppler parameter can be taken as follows:
$$\langle q^{2/3}\rangle\approx 1.2\langle q\rangle^{2/3},\quad
\langle b_{DM}\rangle\approx 2b_0\epsilon,\quad
q\approx 8\tau^2\left({\eta\over 2\epsilon}\right)^{3/2},$$
$$N_b\approx {3\over\sqrt{\pi}b_0\epsilon}\exp(-\xi)
{{\rm erf}(\sqrt{\xi})\over \xi^{1/6} },~~
\xi=\left({\eta\over 2\epsilon}\right)^{3/2}.\eqno(2.9)$$

Relations (2.2), (2.7) and (2.9) link the observed Doppler
parameter with the pancake DM column density, $q$, and indicate
that the distribution function of Doppler parameter is similar
to a gamma distribution. These relations take into account the 
successive merging of earlier formed structure elements -- both 
filaments and pancakes -- what leads, in particular, to
formation of observed galaxy walls at small redshifts.

As was discussed in Paper I, rapid expansion of  pancakes in 
transversal directions decreases $N_{HI}$ below the observational 
limit of $N_{HI}\sim 10^{12}cm^{-2}$. Some fraction of such 
expanded pancakes with smaller $N_{HI}$ can be observed. The 
rapid compression of a pancake along one or both of the 
transversal directions decreases its surface area and also 
the probability to see such a pancake as an absorber.

These inferences are based on theoretical arguments tested
on simulated formation of walls at small redshifts. More
accurate estimates of possible distortions can be obtained
through comparison with representative simulations at high 
redshifts.

As was noted above, we identify $b_{DM}$ as given by (2.6) or 
(2.8) with the {\it observed} Doppler parameter, $b$.

\subsection{Large scale absorber distribution}

The observed spatial distribution of absorbers is very similar 
to the Poissonian distribution. Even so, the large scale
modulation of absorbers distribution is an interesting characteristic.
If absorbers are actually linked to the DM distribution as was
discussed in Sec. 2.2, then it can be expected that this
modulation will be later transformed -- due to the gravitational
instability -- to the galaxy distribution observed at small
redshifts as large and superlarge scale structure.

The large scale matter distribution can be conveniently
characterized by the one dimensional smoothed density field.
The required variance of density can be expressed through the
dimensionless moments of initial power spectrum, $p(k)$. Thus,
for the density smoothed over a scale $r_s$ with a Gaussian
window function we have:
$$\sigma^2_\rho(r_s) = (2\pi)^{-3}\int d^3k~
\exp[-({\bf kr}_s)^2]~p(k)$$
$$= {1\over 4\pi^{3/2}r_s}\int_0^\infty dk~kp(k)~erf(kr_s),
						  	\eqno(2.10)$$
where $k$ is a comoving wave number. For $r_s\rightarrow \infty$
$$\sigma^2_\rho(r_s)\approx {1\over 4\pi^{3/2}r_s}\int_0^\infty
dk~ kp(k) = {\kappa_{cdm}^2\over 4\pi^{3/2}r_s}\int_0^\infty
dk ~kp_{cdm}(k),$$
where $p_{cdm}$ is the standard CDM-like power spectrum with the
Harrison-Zel'dovich asymptotic $p(k)\propto k, as
~~k\rightarrow 0$ and the CDM transfer function and $\kappa_{cdm}$
describes the impact of possible deviations of actual and CDM-like
power spectra used for numerical estimates. For large $r_s$ we have:
$$\sigma_\rho^2(r_s,z)\approx 18\kappa_{cdm}^2\tau^2(z){l_0\over r_s},$$
$$\tau=\tau_\rho\approx 0.24\sqrt{r_s\over l_0}{\sigma_\rho(r_s,z)
\over\kappa_{cdm} }\left(1+{0.1 l_0\over r_s}\right)^{1/6}.
							\eqno(2.11)$$

For such estimates we will use the measured redshift of absorbers
and relations (2.6) and (2.9) to obtain the required DM column 
density of absorbers through the measured Doppler parameter, $b$.

Such approach is similar but not identical to that used by 
Weinberg et al. (1998) and Nusser \& Haehnelt (1998).

\subsection{Properties of gaseous structure elements.}

If the observed Doppler  parameter characterizes the basic
properties of DM pancakes then the hydrogen column density
characterizes the state of the gaseous component trapped by 
the DM pancakes. The gas density and the hydrogen column density
are sensitive to many factors. Firstly, the radiation field
provides the high ionization and bulk heating of hydrogen.
Secondly, the shock compression and heating of gas
accompany the process of merging of richer DM pancakes. The
adiabatic compression or expansion of gas changes also its
temperature and the observed hydrogen column density.

The suitable characteristic of the state of gas is its entropy.
It remains constant during the adiabatic compression and 
expansion of gas and it increases due to all irreversible 
processes such as the shock and bulk heating. The entropy 
decreases only due to the radiative cooling which is usually 
moderate. When gas temperature is rigidly bounded by the 
gravitational potential of DM distribution the entropy and 
the gas density are closely linked.

\subsubsection{Properties of homogeneously distributed hydrogen}

The properties of compressed gas can be suitably related
to  better known parameters of homogeneously distributed gas,
which were described in many papers (see, e.g., Ikeuchi \&
Ostriker 1986). In this case the baryonic density
and the temperature can be taken as
$$\bar{n}_b = 1.2\cdot 10^{-5}\Omega_bh^2(1+z)^3cm^{-3} =
n_0\zeta^3,$$
$$n_0=1.5\cdot 10^{-5}cm^{-3}\Theta_{bar},~~\Theta_{bar}=
\Omega_bh^2/0.02,				\eqno(2.12)$$
$$T_{bg}\approx 1.6\cdot 10^4K,~~~ b_{bg}\approx 16{\rm km/s},~~
\quad \zeta={(1+z)\over 4},$$
and the entropy of the gas can be characterized by the function
$$F_{bg} = T_{bg}/\bar{n}_b^{2/3} = F_0\zeta^{-2},~~ F_0\approx
2~{\rm kev\cdot cm}^2\Theta_{bar}^{-2/3}.
						\eqno(2.13)$$
For reference, the typical entropy of the primordial gas before
reheating, $F_{prm}$, and of gas observed in galaxies, $F_{gal}$,
and in clusters of galaxies, $F_{cl}$, are:
$$F_{prm}\sim 3\cdot 10^{-5}~{\rm kev\cdot cm^2}\approx 1.5\cdot 10^{-5}F_0$$
$$F_{gal}\sim 10^{-3}{\rm kev\cdot cm^2}\approx 10^{-3}F_0$$
$$F_{cl}\sim 300{\rm kev\cdot cm^2}\approx 150 F_0$$
These data illustrate the range of observed variations of
gaseous entropy.

\subsubsection{The hydrogen column density and entropy
of absorbers.}

The observed column density of neutral hydrogen can be written
as an integral over pancake along a line of sight
$$N_{HI} = \int dx {dN_{bar}\over dx}x_H = \langle N_{bar}\rangle
\langle x_H\rangle \kappa.				\eqno(2.14)$$
Here $\langle N_{bar}\rangle$ is the mean column density of
baryons across the absorber, $\langle x_H\rangle$ is the mean
fraction of neutral hydrogen and the dimensionless parameter
$\kappa$ characterizes the nonhomogeneous distribution of neutral
hydrogen across absorber. We will assume that both DM and
gaseous components are compressed together and, so, the column
density of baryons and DM component are approximately proportional
to each other. Therefore, we can take
$$\langle N_{bar}\rangle\approx 
{\bar{n}_bl_0q\over \nu(1+z)}\approx
2.1\cdot 10^{19}cm^{-2}~\xi(b){\tau_z^2\Theta_{bar}\over
\nu\Theta_m},$$
$$\langle \Delta r\rangle\langle\delta_{bar}\rangle=
0.25 l_0q\zeta^{-1}\approx 0.5 \xi(b)\tau_z^2\zeta^{-3}{\rm Mpc}.
							\eqno(2.15)$$
Here $\langle \Delta r\rangle$ and $\langle\delta_{bar}\rangle$
are the mean proper thickness and overdensity of compressed gas
above the mean density, $\xi(b),~\tau_z \& \zeta$ are given by
(2.2) \& (2.5), $\nu = \cos\varphi$ describes the random
orientation of absorbers and the line of sight, and expressions
(2.7) \& (2.9) link functions $q$, $b$ and $\tau$.

Under the assumption of ionization equilibrium of the gas within
a DM pancake and neglecting a possible contribution of macroscopic 
motions to the $b$-parameter ($T\propto b^2$), the fraction of 
neutral hydrogen is
$$\langle x_H\rangle = n_0\zeta^3\langle\delta_{bar}\rangle
{\alpha_{rec}(T)\over \Gamma_\gamma} =
x_0{\langle\delta_{bar}\rangle\over \Gamma_{12} }\zeta^3
\left({b_{bg}\over b}\right)^{3/2},$$
$$\Gamma_\gamma= \Gamma_{12}\cdot 10^{-12}s^{-1},\quad
x_0=4.1\cdot 10^{-6}\Theta_{bar}.$$
The recombination coefficient, $\alpha_{rec}(T)$, and the
photoionization rate due to the extragalactic UV background
radiation, $\Gamma_\gamma$, are taken as
$$\alpha_{rec}(T)\approx 4\cdot 10^{-13}\left({10^4K\over T}
\right)^{3/4} {cm^3\over s},~~\Gamma_{12}\approx 0.7$$
((Black, 1981, Rauch et al. 1997).
Finally, for the column density of neutral hydrogen we have
$$N_{HI} = N_0\xi(b)\left({b_{bg}\over b}\right)^{3/2}
{\langle\delta_{bar}\rangle\over \Gamma_{12}}\kappa
\zeta^3\Theta_H,				\eqno(2.16)$$
$$ N_0 = 8.6\cdot 10^{13}cm^{-2},\quad \Theta_H=
{\Theta_{bar}^2\tau_z^2\over\Theta_m\nu}.$$
and $\xi(b)$ was introduced by (2.2) and (2.6) or (2.9). 

The relation (2.16) links the observed column density of
neutral hydrogen and the Doppler parameter with the column
density of baryons and dark matter, $\xi(b)$, and the degree
of matter compression, $\langle\delta_{bar}\rangle$. The degree
of matter compression depends on the entropy of gas and, so, 
on its evolutionary history that allows us to discriminate 
between various models of absorbers formation and, in particular, 
between adiabatic and shock compressions of the gas accumulated 
within absorbers. Thus, for the adiabatic compression of the gas 
which is more typical for the formation of smaller DM pancakes
$$\langle\delta_{bar}\rangle = (b/b_{bg})^3,\quad
N_{HI}\propto \xi(b)b^{3/2}.			\eqno(2.17)$$
On the other hand, to describe formation of rich pancakes we
can use the function $\xi(b)$ (2.6) or (2.9). In this cases 
we have
$$N_{HI}\approx 0.3N_0{\sqrt{\eta}\over 1+\sqrt{1+\eta^2} }
{\langle\delta_{bar}\rangle\over \Gamma_{12} }\kappa
\zeta^{21/4}\Theta_H,   			\eqno(2.18)$$
$$N_{HI}\approx 0.1N_0\epsilon^{-3/2}{\langle\delta_{bar}
\rangle\over \Gamma_{12} }\kappa\zeta^{21/4}\Theta_H,
						\eqno(2.19)$$
for unrelaxed and relaxed absorbers, respectively. In both
cases the expected correlation between $N_{HI}/\langle
\delta_{bar}\rangle$ and $b$ is negligible and $N_{HI}\propto 
\langle\delta_{bar}\rangle$. For such pancakes the overdensity 
$\delta_{DM}$ obtained in Sec. 2.2.1 is not a good parameter 
because the evolutionary histories and, so, degrees of 
compression of DM and gaseous components are certainly 
different.

As seen from (2.18) \& (2.19), at redshifts $z\geq 3,
~\zeta\geq$ 1 the discussed approach can be applied mainly
to rich absorbers with $N_{HI}\geq 10^{13}$cm$^{-2}$. For 
smaller redshifts, $\zeta\leq$1, this model can also describe 
the properties and evolution of some fraction of poor absorbers 
with $N_{HI}\leq 10^{13}$cm$^{-2}$ connected with DM pancakes. 
The range of application of this model increases for 
$\Gamma_{12}\geq$1, and depends on the local factors which 
can change the parameter $\Theta_H$.

Relations (2.17) and (2.18) allow us to estimate the compression 
factor, $\kappa\langle\delta_{bar}\rangle\Gamma_{12}^{-1}$ and 
the entropy of gas accumulated by absorbers. Two functions
$$F_S = (T/T_{bg})\delta_{bar}^{-2/3}\propto
b^2(\Gamma_{12}/\kappa)^{2/3},~~ \Sigma = \ln F_S,\eqno(2.20)$$
measure this entropy relatively to the entropy assumed for 
the homogeneous intergalactic gas (2.13). These functions
also depend on the photoionization rate, $\Gamma_\gamma$,
and the unknown distribution of neutral hydrogen across
absorbers, described by the factor $\kappa$.

\subsubsection{Adiabatic compression of gas.}

The velocity of infalling gas (2.3) can be rewritten more
transparently as
$$\langle v_{inf}\rangle\approx
62.5km/s~\xi\tau_z^2 \zeta^{-3/2}\Theta_m^{-1/2},	\eqno(2.21)$$
$$\sigma_{inf}\approx 47km/s~\xi^{1/2}\tau_z^2 \zeta^{-3/2}
\Theta_m^{-1/2},\eqno(2.22)$$
and  for poor pancakes with $\xi\leq 0.25\zeta^{3/2}$, when
$\langle v_{inf}\rangle\leq b_{bg}\approx 16km/s$, the formation 
of DM pancakes is accompanied by the adiabatic compression of gas. 
As seen from Eq. (2.17), in this case
$$\xi(b)\propto N_{HI}b^{-3/2}\ll 1,$$
and Eq. (2.2) shows that the homogeneous distribution can be
expected for both $\xi$ and $N_{HI}b^{-3/2}$.

This conclusion can be, in principle, tested using the observed
sample of absorbers. Unfortunately, the available sample of weaker 
Ly-$\alpha$ lines is poor and incomplete and cannot provide the 
statistics required for such test. Moreover, such absorbers are 
more strongly influenced by random local factors that also can 
destroy the expected correlation between $b$ and $N_{HI}$.
On the other hand, the same factors can generate the local
shock waves without any connection with the evolution of
dark matter. Moreover, this sample is complex and it also 
includes absorbers which are not a byproduct of the process 
of DM structure formation.  

\subsubsection{Shock compression of gas.}

Under the shock compression the gas density increases not more
than 4 times while the Doppler parameter increases
proportionally to $v_{inf}$. Simple estimates similar to that
given in Zel'dovich ~\&~Novikov (1983) for "Zel'dovich pancakes" 
and Meiksin (1994) for the collapse of ionized gas into slabs 
show, however, that even for the strong compression of
homogeneously distributed gas the smooth profile of $v_{inf}
\propto\xi$ (2.21) leads to strong adiabatic compression
of the gas before formation of shock waves (see, e.g., Nath 
1997).

For the observed range of Doppler parameters the most plausible
scenario of gas evolution is shock compression of the gas already 
accumulated within the DM confined pancakes and clouds ('grain' 
model). This process is typical for the merging of earlier formed 
structure elements. Such matter distribution reduces the adiabatic 
compression of the gas during the stage when clouds approach each 
other. The shock wave is formed at the moment of merging of clouds 
and only shock compression occurs. This process results in an
essential increase of the entropy because of the limited growth 
of density and larger growth of the temperature $T\propto 
v_{inf}^2$. 

This scenario can be realized if a high matter concentration
within DM confined structure elements has occurred before
creation of observed absorbers. Such strong matter concentration
within clouds, similar in some respects to that considered in the
'mini-halo' model (Rees 1986, Miralda-Escude \& Rees 1993),
is consistent with the theoretical expectations (DD99) because,
for the CDM-like transfer function and Harrison -- Zel'dovich 
primordial power spectrum, the majority of matter should be 
accumulated by low mass clouds with $M_{DM}\sim 10^7M_\odot
(\Omega_m h^2)^{-2}$ already at redshifts $z\geq$ 5. This 
conclusion is consistent with simulations (see, e.g., Zhang 
et al. 1998) which demonstrate that only $\sim$ 5\% of baryons 
remains in a smoothly distributed component. 

The analysis of observational data (see below) shows that for
observed samples there is a strong correlation between
the entropy and the Doppler parameter, while $b$ and $N_{HI}$ are practically
not correlated. For richer
absorbers, with $N_{HI}\geq 10^{13}cm^{-2}$, the entropy
distribution can be approximated as $F_S\propto b^2$ what
agrees well with this model of shock heating of gas with
strongly nonhomogeneous distribution. 

The entropy, $\Sigma$, given by (2.20) is an additive function
which accumulates the successive contributions of shock and
bulk heating during all evolutionary history of a given
gaseous element. If the shock heating of gas dominates, then
the growth of entropy, $\Sigma$, can be described as a random
process -- similar to the Brownian motion -- with a successive
uncorrelated jumps of entropy at each step. This means that
the expected PDF of the entropy, $\Sigma$, should be similar 
to Gaussian. 

For the samples of poorer absorbers with $N_{HI}\leq 10^{13}
cm^{-2}$ there are both substantial correlations between $b$ 
and $N_{HI}$ and between the entropy and the Doppler parameter. 
This fact indicates the complex character of such absorbers. If 
they can be associated with pancake-like DM elements then for 
some fraction of such absorbers the adiabatic compression or 
bulk heating could be significant, while some of them could 
be formed due to expansion of richer earlier formed pancakes. 
Some fraction of such absorbers can be also related to artificial 
pancakes discussed by McGill (1990) and Levshakov \& Kegel 
(1996, 1997) and to absorbers situated within "minivoids" what 
increases uncertainty in their discrimination. 

\subsubsection{Bulk heating and cooling of the gas.}

The bulk heating and cooling of the compressed gas can be also 
essential for the evolution of properties of  absorbers. The main 
factor is the random spatial variation of the spectrum of ionizing 
UV radiation field generated by local sources (Zuo 1992, Fardal~\&~ 
Shull 1993). The action of this factor can be characterized by the 
mean energy injected at photoionization, $T_\gamma\approx$(5 -- 10)
$\cdot 10^4K$ for suitable spectra of UV radiation (Black 1981).
In extremal cases such variations are observed as a proximity
effect (Bajtlik, Duncan \& Ostriker 1988). In the model of DM
confined absorbers discussed here, when the gas temperature is
given by the gravitational potential of DM component, the bulk
heating changes the density and entropy of compressed gas.

The influence of the bulk heating can be enhanced by the
pancake disruption due to the clustering of both DM and
baryonic components. This process is usually accompanied
by adiabatic reduction of temperature in expanded regions 
when the role of the bulk heating becomes more important. 
As is shown in Appendix B, in the general case, for redshift 
dependent $T(z)~\&~T_\gamma(z)$ we have
$$F_S^{3/2}(z)=F_S^{3/2}(z_f)+\alpha_s\int_z^{z_f}dx
{H_0\over H(x)}{T_\gamma(x)-T(x)\over (1+x)T_{bg}},$$
$$\alpha_s = 1.5h^{-1}\zeta^3\Theta_{bar},		\eqno(2.23)$$
where the entropy function $F_S$ was introduced by (2.20)
and $z_f$ is the redshift of pancake formation.
For example, under condition of slow variation of the
DM distribution when $T(z)\approx {\rm const.}$ and for
$T_\gamma\approx {\rm const.}$ we have
$$F_S^{3/2}(z)=F_S^{3/2}(z_f)+\alpha_s~H_0[t(z)-t(z_f)]
{T_\gamma - T\over  T_{bg} },$$
$$t(z)=\int_z^\infty {dx\over H(x)(1+x)},$$
that indicates the slow growth of entropy and drop of the
density of absorbers. 

At higher redshifts the proper separation of structure
elements is about  $D_{sep}\sim$0.5 -- 1$h^{-1}$Mpc that
is comparable with sizes of galactic halos observed at
small redshifts  $\sim 0.2h^{-1}$Mpc (see, e.g., Bahcall
et al. 1996). This means that the evolution of structure
elements at such redshifts could be interdependent
and activities in galaxies can increase the gas entropy within
nearby pancakes. Analysis of Cen~\&~ Ostriker (1993) shows
that the input of explosive energy can be essential in the 
vicinity of virialized objects. In this case the gas 
overdensity decreases in proportion to the injected
energy and the additional correlations between the Doppler
parameter and column density of neutral hydrogen do not
appear. The observations of metal systems together with
relatively weak Ly-$\alpha$ lines shows that sometimes the
influence of such heating can be important.

\subsubsection{Evolution of hydrogen column density}

The observed column density of absorbers is changed
due to irreversible processes such as shocks and bulk
heating, and due to adiabatic expansion and/or compression
of pancakes caused by the transversal motions of DM component
(DD99, Paper I). The heating of the gas does not distort the
DM distribution and the ionization equilibrium and, as was
discussed in Sec. 2.4.5, also the relatively slow decrease of 
$N_{HI}\propto \langle\delta_{bar}^2\rangle$. The adiabatic
compression and expansion of DM and gaseous components also 
does not distort the ionization equilibrium, but changes the 
observed column density $N_{HI}\propto b^{9/2}$, and can 
lead to the appearance of rare absorbers with larger and 
smaller $N_{HI}$ and $b$.

These simple arguments demonstrate the important role of
adiabatic evolution of DM pancakes for the formation of
observed sample of absorbers. Together with the merging 
these processes lead to the fast evolution of linear density 
of absorbers discussed in Paper I and are essential for the 
formation of both rare absorbers, with larger $N_{HI}$, and 
weaker absorbers. 

\subsection{Theoretical expectations}

The previous consideration can be shortly summarized as follows:
\begin{enumerate}
\item{} For the DM confined absorbers the observed Doppler
parameter, $b$, is closely linked to the column density of DM
pancakes, $q$, and the expected PDFs of the $b$ parameter, (2.7)
and (2.9), are similar to the gamma-distribution. The density
field can be described as a set of discrete pancakes with masses 
$\propto q(b)$.
\item{} For such absorbers only weak correlation between the
Doppler parameter, $b$, and the column density of neutral hydrogen,
$N_{HI}$, is expected under the assumption of ionization equilibrium.
\item{} The same factors result in the strong correlation
between the entropy of compressed gas and the Doppler parameter. 
Due to merging and shock heating the gas entropy increases. This 
growth  can be described as a random process with the function 
$\Sigma = \ln F_S$ increasing discontinuously with jumps depending 
on the mass of merging pancakes. The possible bulk heating caused 
by the local sources results also in random jumps of entropy. More 
detailed description of entropy evolution can be obtained on 
the basis of Fokker-Plank equation that implies however more 
detailed description of the pancake evolution and assumptions
about the local activity of galaxies. In any case approximately 
Gaussian distribution of $\Sigma$ can be expected.
\item{} As is seen from equations (2.18), (2.19) \& (2.20) 
for such pancakes the approximately Gaussian PDFs are also expected 
for $\log\delta_{bar}$ and $\log N_{HI}$.
\item{} The adiabatic compression of the gas dominates for
poorer DM pancakes with $N_{HI}\leq 10^{13}cm^{-2}$ and creates
much stronger correlation between $N_{HI}$ and $b$. This
correlation can be, however, partly destroyed by the action
of local random factors. Formation of such absorbers is often 
not accompanied by formation of DM structure elements. 

\end{enumerate}

As is seen from (2.16) and (2.20) the {\it observed} estimates
of entropy, $\Sigma$, and the column density, $N_{HI}$, depend 
on the intensity of UV ionizing radiation (factor $\Gamma_{12}$). 
This means that local variations of the ionizing rate introduce
the essential random factor in observational estimates of both
entropy and column density. The rapid growth of observed number
of galaxies at $z\leq 3$ (Steidel et al. 1998; Giavalisco et al.
1998) correlates well with the  observed rapid evolution of linear
density of absorbers, $n_{abs}$, (Paper I) what points toward
galaxies as the essential factor of observed evolution of
absorbers.

\subsection{Characteristics of absorbers in simulations}

High resolution simulations of dynamical and thermal evolution 
of gaseous component cited above give us valuable examples of 
absorbers formed at redshifts $z\leq$5 and allow to clarify many 
peculiarities and properties of these processes. In particular, 
they confirm the existence of high density clumps, filamentary 
and sheet-like components of structure at high redshifts and 
demonstrate that these elements accumulate up to 95\% of 
baryons and reproduce well the main observed characteristics 
of absorbers. They confirm the domination of adiabatic 
compression of baryonic component in less massive pancakes 
and show that some fraction of weaker absorbers can be identified 
with underdense structures (Bi \& Davidsen 1997; Zhang et al. 
1998; Dav\'e et al. 1999). List of these examples can be 
essentially extended.    

At the same time, the abilities of simulations in statistical 
description of absorbers evolution are limited as the main 
simulations are performed with small box sizes ($\sim$10 Mpc) 
that is comparable with the mean separation of galaxy filaments 
at small redshifts, and is smaller then the mean separation of 
richer galaxy walls ($\sim$40 -- 60$h^{-1}$Mpc) and even their 
Lagrangian size ($\sim$15 -- 20$h^{-1}$Mpc). The small box sizes 
restrict the important influence of large scale perturbations 
on the evolution of small scale structure and does not allow 
to establish direct connection of structure at high redshifts 
with galaxy distribution observed at small redshifts. 

Moreover, some results obtained in simulations should be explained 
in more details. Thus, for example, in all papers the very important 
problem of quantitative characteristics of absorbers morphology 
is not discussed and the merging of pancakes and filaments is not 
considered. Such merging is certainly responsible for the formation 
of observed galactic walls and was discussed in DD99, DDMT, Paper I 
and in Sec. 2.2 \& 2.3 above as one of the main factors of absorbers 
evolution. Its action can be easily traced using the analysis of 
entropy of compressed gas. The distribution function of temperature 
(Fig. 10 in Zhang et al. 1998) is quite different from the observed 
distribution of Doppler parameter. The strong contribution of 
macroscopic motions in simulated Doppler parameter plotted in Fig. 
13 increases it about of 2 -- 5 times with respect to the thermal 
value what probably implies the supersonic character of typical 
macroscopic motions within simulated structure elements. In contrast, 
the analysis of DM evolution in large boxes demonstrates an approximate 
isotropy of subsonic velocity dispersion within DM structure elements 
at small redshifts, and its close connection with the measured 
richness of such element (DMRT, DDMT). The physical reasons of 
the power distribution of $N_{HI}$ plotted in Fig. 15 of the same 
paper, in the range of 5 orders of magnitude, are also not explained. 

Even these examples demonstrate an essential scatter of some 
quantitative characteristics of structure obtained with different 
codes and parameters of simulations (see also discussion in Melotte 
et al. 1997; Splinter et al. 1998; Theuns et al. 1999). They show 
also the usefulness of more detailed comparisons and tests of
consistency  of results obtained with different approaches and box
sizes, and  based on larger set of characteristics. Bearing in mind
the relatively small number of such complicated simulations, these
differences, list of which can also be continued, seem to be natural.

\section{ The database.}

\begin{table}
\caption{QSO spectra from the literature}
\label{tbl1}
\begin{tabular}{cccccc}
Name&$z_{em}$&$z_{min}$&$z_{max}$&FWHM&No \\
               &     &   &   &km/s&of lines\\
$0000-260^{1}$ & 4.11&3.4&4.1&~~7& 431\\
$0014+813^{2}$ & 3.41&2.7&3.2&~~8& 262\\
$0956+122^{2}$ & 3.30&2.6&3.1&~~8& 256\\
$0302-003^{2}$ & 3.29&2.6&3.1&~~8& 266\\
$0636+680^{2}$ & 3.17&2.5&3.0&~~8& 313\\
$1946+766^{3}$ & 3.02&2.4&3.0&~~8& 461\\
\hline
$0055-259^{4}$ & 3.66&2.9&3.1& 14& 313\\
$2126-158^{5}$ & 3.26&2.9&3.2& 11& 130\\
$1700+642^{6}$ & 2.72&2.1&2.7& 15&~~85\\
$1225+317^{7}$ & 2.20&1.7&2.2& 18& 159\\
$1101-264^{8}$ & 2.15&1.8&2.1&~~9&~~84\\
$1331+170^{9}$ & 2.10&1.7&2.1& 18&~~69\\
\hline
$1033-033^{10}$& 4.50&3.7&4.4& 18& 299\\
$2206-199^{11}$& 2.56&2.1&2.6& 11& 101\\
\vspace{0.15cm}
\end{tabular}

1. Lu et al. (1996),
2. Hu et al., (1995),
3. Kirkman \& Tytler (1997),
4. Cristiani et al. (1995),
5. Giallongo et al. (1993),
6. Rodriguez et al. (1995),
7. Khare et al. (1997),
8. Carswell et al. (1991),
9. Kulkarni et al. (1996),
10. Williger et al. (1994),
11. Rauch et al. (1993),
\end{table}

The present analysis is based on the spectra available in the
literature. The list of the used sources of data is given in Table 1.
As was discussed in Paper I absorbers with $b\geq$ 17km/s or
$\log(N_{HI})\leq$ 14 can be probably related to the sheet-like
component of structure. In spite of statistical character of such
discrimination it allows us to obtain more homogeneous sample
of sheet-like absorbers used for comparison with theoretical expectations.
The list of available Ly-$\alpha$ lines was arranged
into three samples. Two of them, $Q_{12}$ and $Q_{14}$ include 2073
and 2378 lines from the first 12 and all 14 spectra, respectively,
under conditions that $b\geq$ 17km/s and $10^{13}cm^{-2}\leq N_{HI}
\leq 10^{14} cm^{-2}$. First condition allows us to discriminate 
and to exclude absorbers which could be possibly formed due to 
the adiabatic compression.

The sample $Q_{0612}$ contains 469 lines with $N_{HI}\leq 10^{13}
cm^{-2}$ and $b\leq$ 30km/s from the first 6 QSOs. This sample is
incomplete as was discussed by Hu et al. (1995). It can be used
for the estimates of properties of poor absorbers.

The line distribution over redshift is nonhomogeneous and the majority
of lines are concentrated at $z\approx$ 3. The distribution of
absorbers
at $z\geq 3.2$ is based mainly on the spectrum of QSO
0000-260 (Lu et al. 1996) and here the line statistics is insufficient.
The inclusion of the spectrum of QSO 1033-033 extends the redshift
interval up to $z\approx$ 4.4, but cannot eliminate small
representativity of the sample at such redshifts.

\begin{table*}
\begin{minipage}{130mm}
\caption{Main parameters of Ly-$\alpha$ absorbers.}
\label{tbl2}
\begin{tabular}{ccc ccc ccc ccc c } 
     &$z_{min}$&$z_{max}$&$N_{line}$&$\zeta$&$\langle b\rangle$&$\langle
q\rangle$&$c_2$&$\tau_b$&$r_{bH}$&$\langle F_S\rangle$&$p_{bs}$&
$\sigma_S$\cr
&&&&&km/s&&&&&&&\cr
\hline
\multicolumn{13}{c}{$Q_{12}$}\cr
   a     &1.7  &4.1 &2073&0.98&36.9&0.2&1.6&0.17&~0.04&4.8&1.89&0.9\cr
   b     &2.0  &3.0 &1118&0.92&37.6&0.2&1.5&0.17&-0.04&5.0&2.14&0.9\cr
   c     &2.5  &3.5 &1392&0.98&36.8&0.2&1.9&0.18&~0.02&4.9&2.05&0.9\cr
   d     &3.0  &4.1&~~813&1.10&36.0&0.2&2.0&0.19&~0.16&4.8&1.59&0.8\cr
\hline
\multicolumn{13}{c}{$Q_{14}$}\cr
         &1.7  &4.4 & 2378&1.00&38.2&0.2&1.9&0.16&~0.11&6.1&1.91&0.8\cr
         &2.0  &3.0 & 1190&0.92&37.3&0.2&1.6&0.17&-0.02&4.8&2.10&0.8\cr
         &2.5  &3.5 & 1403&0.98&36.8&0.2&1.7&0.17&-0.02&4.6&2.05&0.8\cr
         &3.0  &4.4 & 1046&1.11&39.5&0.3&1.8&0.21&~0.23&5.4&1.86&0.8\cr
\hline
\end{tabular}

The correlation coefficient $r_{bH}$, exponent $p_{bs}$ and the entropy 
dispersion $\sigma_S$ are introduced in Secs. 4.3 and 4.2, respectively.
\end{minipage}
\end{table*}

\begin{figure}
\centering
\epsfxsize=7 cm
\epsfbox{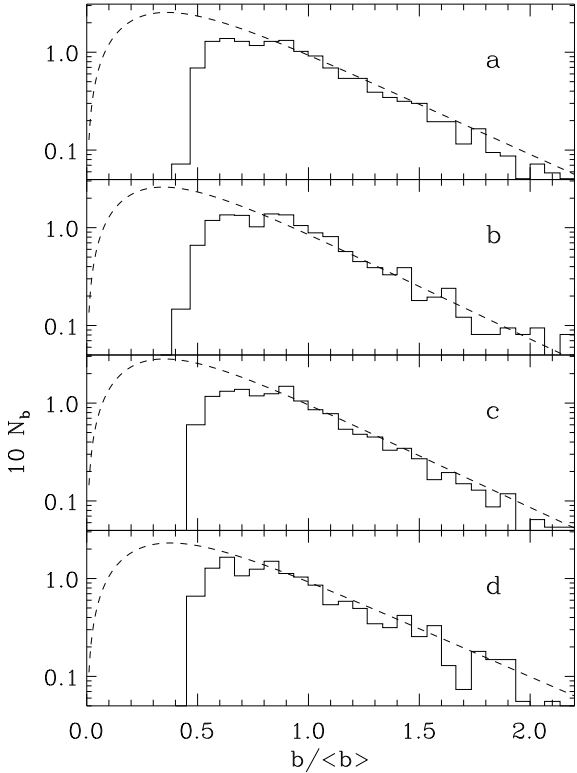}
\vspace{0.7cm}
\caption{The distributions of Doppler parameter,
$b/\langle b\rangle$, for the sample $Q_{12}$ of Ly-$\alpha$
lines for different redshifts. The best fits (4.1) are plotted
by dashed lines. The main parameters of subsamples
are listed in Table 2.}
\end{figure}

\begin{figure}
\centering
\epsfxsize=7 cm
\epsfbox{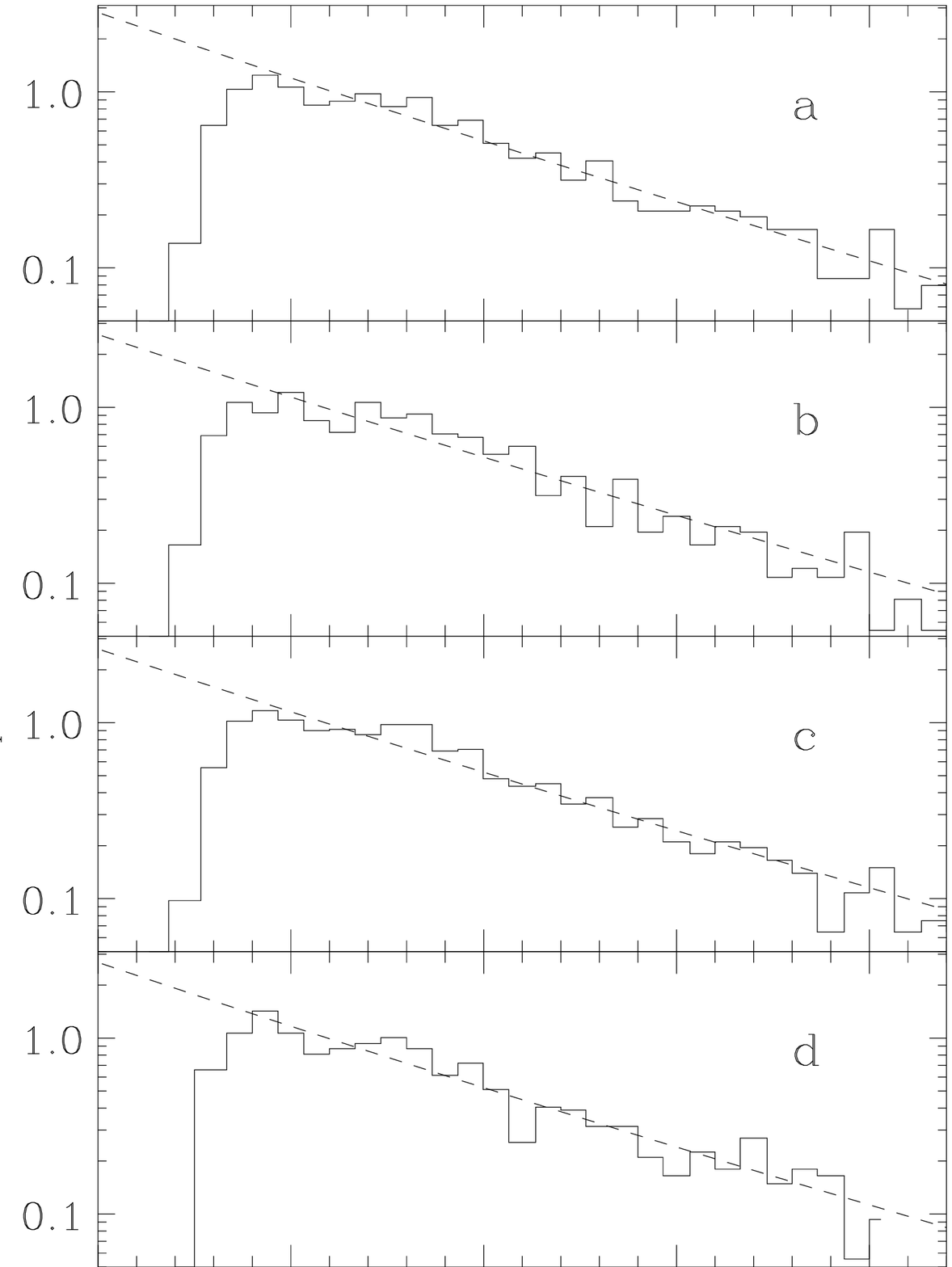}
\vspace{0.7cm}
\caption{The distributions of DM column density,
$q/\langle q\rangle$, for the sample $Q_{12}$ of Ly-$\alpha$
lines for different redshifts. The best fits (2.2) are plotted
by dashed lines. The main parameters of subsamples
are listed in Table 2.}
\end{figure}

\section{Statistical characteristics of absorbers}

Some statistical characteristics of absorbers were found for
samples $Q_{12}$ and $Q_{14}$ for four ranges of redshifts.
Main results are listed in Table 2 and plotted in Figs. 1 -- 4.

\subsection{Distribution of observed Doppler parameter, $b$, and
DM column density, $q$}

As was discussed in Sec. 2.2, the observed distribution of
Doppler parameter, $b$, can be interpreted as the distribution
of DM column density of absorbers. To check this hypothesis the
observed distribution of $b/\langle b\rangle$ was fitted to the
two parameters function similar to (2.6) \& (2.7)
$$N_b = c_1{\rm erf}(\sqrt{y}){\sqrt{2+y}\over 1+y}\exp(-y),\eqno(4.1)$$
$$y=\sqrt{1+c_2^2b^2/\langle b\rangle^2}-1 = 
{c_2^2b^2/\langle b\rangle^2\over 1+\sqrt{1+c_2^2b^2/\langle 
b\rangle^2}},$$
for both samples and four different ranges of redshifts.
For each absorber the dimensionless column density of DM component,
$q$, was found using Eq. (2.6) and its distribution was fitted to
a two parametric function
$$N_q\approx c_3 e^{-y}~{{\rm erf}
(\sqrt{y})\over\sqrt{y}} ,\quad y=c_4 q/\langle q\rangle.\eqno(4.2)$$
The main results are listed in Table 2 and plotted in Figs. 1 \& 2.

The observed properties of the Doppler parameter, $b$, are found
to be surprisingly stable and independent of samples or redshift
ranges. In all cases we have
$$\langle b\rangle = 37km/s,\quad \sigma_b =0.55\langle b\rangle.
							\eqno(4.3)$$

The functions (4.1) \& (4.2) fit well the observed $b \& q$
distributions for all samples under consideration. Fit parameters
$c_1 \& c_2$ and $c_3 \& c_4$ describe the cutoff in observed PDFs
at $b\approx 0.5\langle b\rangle$. This cutoff naturally appears
for the Doppler parameter of gaseous component as in both samples, 
$Q_{12}~\&~Q_{14}$, absorbers with $b\leq 17km/s\approx 0.5\langle 
b\rangle$ were excluded from the analysis. This cutoff
increases the observed $\langle b\rangle$ and $\langle q\rangle$
by the factor of $c_2$ and $c_4$, respectively, in comparison
with theoretical expectations for the velocity and column density
of DM component. Some deficit of absorbers with larger
$b\geq$ (2 -- 2.5)$\langle b\rangle$ is seen in Fig. 1.
For the PDF (4.1) the expected value is $\sigma_b=0.7\langle
b\rangle$. The difference between observed and expected
$\sigma_b$ is also explained by the same cutoff and for
$y\geq$ 0.5, the expected $\sigma_b=0.6\langle b\rangle$ is
practically identical to (4.3).

\begin{figure}
\centering
\epsfxsize=7 cm
\epsfbox{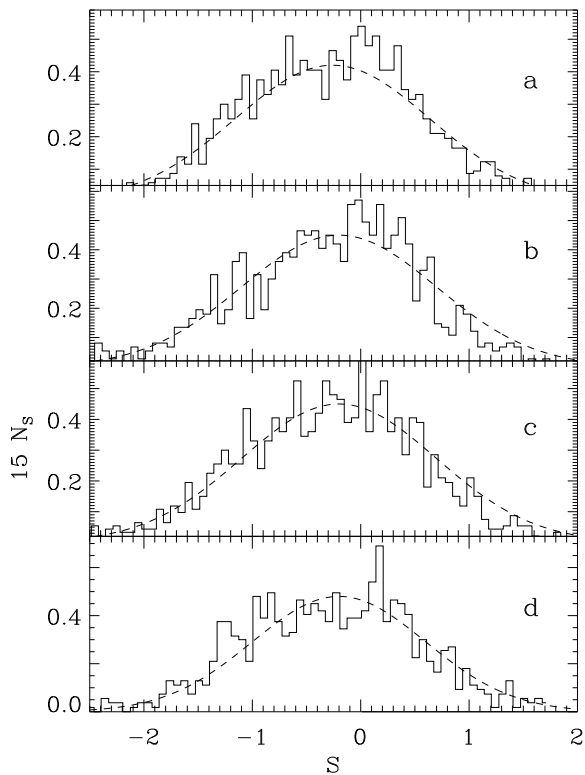}
\vspace{0.7cm}
\caption{The distribution function of absorbers 'entropy',
$S$, for the sample $Q_{12}$ for four different redshifts.
The best Gaussian fits are shown by dashed lines.
}
\end{figure}

The fitting parameters $c_2 \& c_4$ allow us to correct measured
$\langle b\rangle$ and $\langle q\rangle$ for the cutoff at
$b\approx 0.5\langle b\rangle$ and to estimate the amplitude and
time scale of the DM structure evolution, $\tau$, as given by 
(2.2) and (2.6):
$$\tau_b = (1+z)\tau = (1+z)\sqrt{\langle q\rangle\over 6.55 c_4}$$
$$ = (1+z)\sqrt{\langle b\rangle\over 720 km/s ~c_2}\zeta^{3/4}
\Theta_m^{1/4}\approx (0.17\pm 0.1).			\eqno(4.4)$$
$c_2$ and $\tau_b$ listed in Table 2 are found to be weakly dependent
on redshift $z$ and on subsamples used. This value $\tau_b$ is roughly 
consistent with the values $(1+z)\tau\sim$0.2 -- 0.4 expected for 
low density cosmological models (Appendix A). 

We can also estimate the typical proper thickness of DM absorbers,
$h_{DM}$, which is linked with $\tau$ by Eq. (2.3), we have
$$\langle h_{DM}\rangle\approx {8\over\sqrt{\pi}} l_0{\tau^2\over 1+z}
\sim {130{\rm kpc}\over\Theta_m \zeta^3}\left({\tau_b\over
0.17}\right)^2.						\eqno(4.5)$$
This value is approximately consistent with sizes found by Dinshaw
et al. (1995) and is similar to the measured sizes of galactic halos
at small redshifts (Lanzetta et al. 1995; Bahcall et al. 1996) and
to results obtained in Paper I. We cannot obtain reasonable estimates
of the size, $\langle\Delta r \rangle$, and overdensity
$\langle\delta_{bar}\rangle$, of gaseous halo because they depend on
the unknown parameter $\kappa$ introduced in (2.14) which characterizes 
the distribution of neutral hydrogen across absorber. 

\subsection{Entropy of absorbers.}

The observed characteristics of the entropy described by the
functions $F_S$ and $\Sigma$ (2.20) were found for the same
subsamples of absorbers and the main results are plotted in
Figs. 3 and listed in Table 2. The function $F_S$ can be
fitted to the expression
$$F_S = \langle F_S\rangle\left({b\over\langle b\rangle}\right)
^{p_{bs}}\exp(S),~ S = \Sigma-p_{bs}ln\left({b\over\langle 
b\rangle}\right),  \eqno(4.6)$$
which discriminates between the regular variations of the
entropy described by the first terms in (4.6), and the integral
action of random factors described by the function $S$. The
PDF of $S$ plotted in Figs 3 is fitted to Gaussian functions with
$\sigma_S$ listed in Table 2. It is found to be sufficiently
stable and weakly sample depended. At larger $S$ the deficit of
absorbers with higher entropy is seen as an asymmetric shape of
PDF. It can be caused by the deficit of observed absorbers with
larger $b$ and smaller $N_{HI}$. Estimates of $\langle F_S\rangle$
listed in Table 2 show that the entropy of absorbers with larger
$b\geq (4 - 5)\langle b\rangle$ is similar to the entropy of gas
accumulated by clusters of galaxies as is given in Sec. 2.4.1.

The value $p_{bs}\approx$2 shows that merging of pancakes is
the main factor of entropy evolution. It shows also that among 
pancake merging there is one which provides the main jump in
both the Doppler parameter and entropy and their correlation. 
The impact of smaller jumps in the course of the merging 
of pancakes, random local variations of ionizing UV radiation, 
a possible bulk heating and other random factors is well described 
by the Gaussian distribution of S-function. It agrees well with 
negligible correlation of measured $N_{HI}$ and $b$, since the 
correlation coefficient $r_{bH}\leq$0.1.  

\begin{figure}
\centering
\epsfxsize=7 cm
\epsfbox{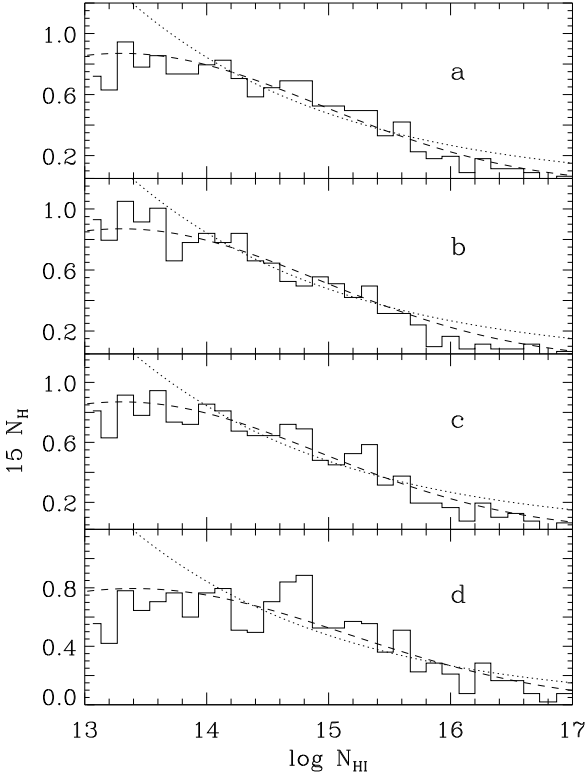}
\vspace{0.7cm}
\caption{The distribution function of $\log N_{HI}$ for the
sample $Q_{12}$ for four ranges of redshifts. Parameters of
samples and fitting parameters are listed in Table 2.
The best fits (4.8) and (4.9) are plotted by dashed and dotted
lines, respectively.
}
\end{figure}

The small range of observed redshifts, 0.8$\leq \zeta\leq$1.2, 
does not allow more detailed analysis of possible variations 
of the mean entropy with $z$. Nonetheless, the weak decrease of 
$p_{bs}$ and corresponding growth of $r_{bH}$ at $z\geq$ 3 shows
that for larger $z$ the merging of pancakes could be accompanied
by other irreversible processes and/or by adiabatic gas compression.

\subsection{Observed distribution of column density of HI.}

One of the most enigmatic feature of the distribution of
observed column density of neutral hydrogen
is that it can be approximated by a single power law
with a power index $\beta_H=1.5\pm 0.05$ in the range
$10^{13}cm^{-2}<N_{HI}<10^{22}cm^{-2}$ (see, e.g., Tytler 1987,
Hu et al. 1995, Kim et al. 1997). For smaller $N_{HI}$ the
deviations from the fit are usually assigned to the
incompleteness of samples. It is interesting that similar 
power distribution was found (but not explained) in simulations 
as well (Zhang et al. 1997, 1998) 

In the considered model of absorbers, when the gas temperature and
the Doppler parameter, $b$, are hardly linked with characteristics
of DM distribution, the well known negligible correlation
of $N_{HI}$ and $b$ indicates the weak connection of $N_{HI}$
with the properties of DM component of absorbers, such as, in 
particular, the DM surface density of pancake, $q$. For samples
under consideration the correlation coefficient
$$r_{bH} = {\langle log N_{HI}*
log~b\rangle - \langle log N_{HI}\rangle\langle log~b\rangle
\over \sigma_{log N_{HI}}\sigma_{log~b} }, \eqno(4.7)$$
was estimated directly using measured $b$ and $log N_{HI}$ and 
in all the cases $r_{bH}\leq$0.1 were found (Table 2). For the 
DM dominated absorbers the column densities of baryonic component 
and, therefore, also $N_{HI}$ depend mainly on the entropy of 
gas, $F_S$ or $\Sigma$, which accumulates the contribution 
of irreversible processes during all evolutionary history 
of the compressed gas. The action of this factor disconnects 
the surface densities of baryonic and DM components, $N_{HI}$ 
and $q \& b$. Essential variations of the entropy implies 
essential variations of observed $N_{HI}$ even for the same 
temperature, $b$, and DM surface density, $q$. This means 
that it is difficult, if not impossible, to explain this joint 
power distribution in such a wide range of column density.

The negligible correlation between the observed $b$ and $N_{HI}$
implies also that 
$$N_{HI}\propto \langle\delta_{bar}\rangle\propto (F_S/b^2)^{-2/3},
~~\log N_{HI}\approx -{2\over 3}S + {\rm const}.$$
This means that at least for the considered subsample of 
pancake-like absorbers the PDF for observed $N_{HI}$ should 
be similar to  PDF of the function $S$, discussed in previous 
Subsec. Therefore, the expected PDF of $\log(N_{HI}/N_m),
\quad N_m =10^{13}cm^{-2}$, is Gaussian and can be written as
$$N_H = N_0\exp\left[-{1\over 2\sigma_H^2}\log^2
\left({N_{HI}\over N_m}\right)\right].		\eqno(4.8)$$

The distribution function of $\log(N_{HI})$ is plotted in Fig. 
4 for four redshift intervals, for the sample of 12 QSO and
for all observed lines, with $N_{HI}\geq 10^{13}cm^{-2}$. The 
physical model of absorbers evolution introduced in Secs. 2.2 
and 2.4 applies only to absorbers with $N_{HI}\leq 10^{14}
cm^{-2}$. The distribution of stronger absorbers, which are 
probably linked with filaments and high density clumps, should 
be discussed in a context of spatial distribution and evolution 
of these components of structure. The statistics of absorbers 
with $N_{HI}\geq 10^{14}cm^{-2}$ is limited, and the right part 
of Figs. 4 is rather  illustrative but, even so, these Figs. 
demonstrate that the relation (4.8) also approximately fits the 
observed distribution of $N_{HI}$ up to $N_{HI}\sim 10^{17}cm^{-2}$. 

The power distribution of $N_{HI}$ corresponds to the exponential
distribution of $log(N_{HI})$ and can be written as
$$N_H\propto \exp\left[2.30(1-\beta_H)\log
\left({N_{HI}\over N_m}\right)\right].		\eqno(4.9)$$
This fit is plotted in Figs. 4 by dotted lines for $\beta_H\approx$
1.5 (Hu et al. 1995; Kim et al. 1997). It agrees with the observed
PDF $N_H$ for larger $log(N_{HI})$ but predicts some excess of
weaker lines.

These results show that the problem deserves further investigation
in a wider range of redshifts with a more representative sample of
absorption lines.

\section{Properties of weaker absorbers}

The sample of observed weaker absorbers with $N_{HI}\leq
10^{13}cm^{-2}$ and $b\leq$ 30km/s is incomplete and composed
of only 469 lines in 6 QSOs listed in Table 1. This factor 
as well as the much more complicated composition of this 
subpopulation indicate illustrative character of our 
consideration in this subsection.

As was discussed in Sec. 2.2 such absorbers can be created by 
adiabatic compression of homogeneous gas or the
disruption and adiabatic expansion of earlier formed rich
pancakes. For the first subpopulation the entropy is not
changed and $F_S\approx$ 1 can be expected. For such pancakes
$b$ and $F_S$ are evidently uncorrelated, but significant
correlation between $N_{HI}$ and $b$ can be expected. For
the second subpopulation $F_S\geq$ 1 and, as before, it correlates
with $b$, but $N_{HI}$ and $b$ are only weakly correlated.
Moreover, for such absorbers the contribution of 
artificial pancakes discussed in McGill (1990) and Levshakov 
\& Kegel (1996, 1997) and absorbers formed within "minivoids" 
(Zhang et al. 1998) can be more significant.  

This means that the population of weaker absorbers consists of 
objects with different evolutionary history, and the available
observational data do not allow us to discriminate between these
subpopulations. So, we have to restrict our
consideration to the analysis of PDF for the column density of DM
component $q\propto N_{HI}b^{-3/2}$, as was described in Sec.
2.4.3.

This PDF, $N_q$, is plotted in Fig. 5. As was expected
for smaller $q$, $N_q\approx {\rm const}.$ especially
at larger redshifts $z\geq$ 3. The significant correlation
between $\log N_{HI}$ and $\log b$, since $r_{bH}\approx$ 0.4,
agrees also with the probable substantial contribution of
adiabatically compressed subpopulation of weaker absorbers.
The 'tail' of absorbers with larger $q/\langle q\rangle$ can be
associated with  the subpopulation of disrupted and/or expanded
earlier formed richer absorbers. The fraction of such absorbers
increases for smaller $z$.

\section{Spatial distribution of absorbers}

The observed redshift distribution of absorbers contains
significant information about the spatial matter
distribution at high redshifts, as suggested in particular
by Oort (1981, 1984). The available absorption spectra cover
typically the range $D\sim$ 200 -- 300$(\Omega_m)^{-1/2}
h^{-1}$Mpc, whereas a mean separation of weak absorbers with
$N_{HI}\approx 10^{12}cm^{-2}$ is $\sim$1$(\Omega_m)^{-1/2}
h^{-1}$Mpc. This allows us to analyze both the
small and large scale matter distribution.

\subsection{Small scale absorber distribution}

During last years  weak clustering of Ly-$\alpha$ absorbers
on small scales ($\Delta v\leq$ 300km/s) have been found for a few
quasars (Webb 1987; Cristiani et al. 1995; Hu et al. 1995; Ulmer
1996; Fernandes-Soto 1996). For many other objects such clustering
is negligible and the absorbers distribution is nearly Poissonian.
Stronger small scale correlation, found for metal lines, is
naturally explained since several lines can be generated in
the same gaseous cloud.
At the same time much stronger correlation of galaxies is
found both at small and even high redshifts (see, e.g.,
discussion in Steidel et al. 1998; Governato et al. 1998; 
Giavalisco et al. 1998)).

\begin{figure}
\centering
\epsfxsize=7 cm
\epsfbox{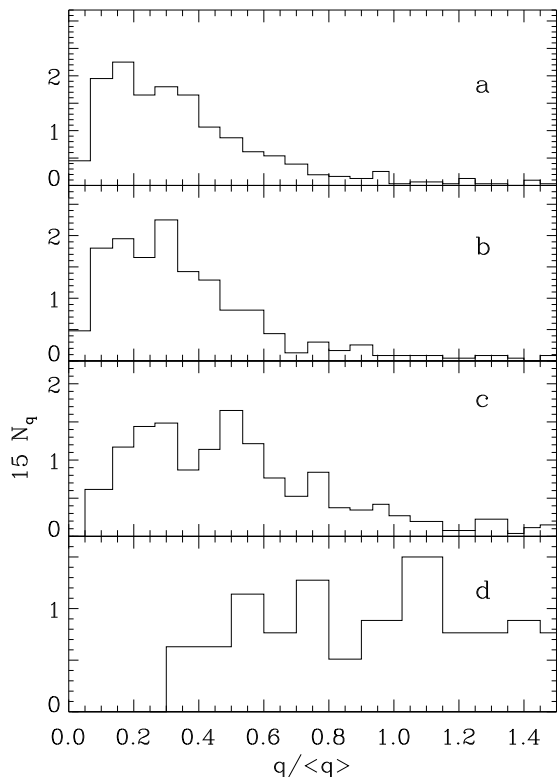}
\vspace{0.7cm}
\caption{The distributions function of $q\propto N_{HI}b^{-3/2}$
for the sample of weaker absorbers with $N_{HI}\leq 10^{13}cm^{-2}$
and $b\leq$30 km/s for the same ranges of redshifts.
}
\end{figure}

This divergence can be naturally explained in the framework
of discussed here approach, when absorbers are associated with
structure elements -- filaments and pancakes -- rather then
with galaxies. The distribution of structure elements along
a random straight line is expected to be Poissonian - like
(DD99, DDMT) what is consistent with the observed distribution 
of filaments and walls at small redshifts (Doroshkevich et al.
1996). In contrast, the point - like high density clumps, which 
can be associated with 'galaxies', are mainly incorporated 
into filaments and massive pancakes. Even for randomly
distributed filaments and pancakes this concentration
introduces some regularity in the spatial distribution
of such clumps in comparison with the 3D Poissonian
distribution. As was shown by van de Weygaert (1991) and
Buryak \& Doroshkevich (1996) such point concentration
generates the 3D correlation function similar to that
observed for galaxies.

Similar situation occurs, for example, for the LCRS,
where the usual correlation function of galaxies was found 
(Tucker et al. 1997). However, for the same observed sample
the 1D distribution of both filaments and walls (in radial and
transversal directions) were found to be Poissonian -- like.

\subsection{ Large scale modulation of absorbers distribution}

The smoothed absorbers distribution can be compared  
with expectations discussed in Sec. 2.3. Three expressions 
for DM distribution are examined: 1) as was discussed
in Sec. 2.2.2, for relaxed pancakes $q\propto b^{3/2}$ can be
expected, 2) if the compressed matter is not completely relaxed,
then $q\propto \sqrt{1+\eta^2}-1$ is more probable, and 3) for
comparison with the previous definition, the hypothesis 
$q\propto N_{HI}$ is also tested. The third expression is 
correct for weaker absorbers discussed in Sec. 2.4.3 and Sec. 
5 and, possibly, can be also applied to high density peaks 
associated with galaxies but its application to main part of 
absorbers is in question.

\begin{figure}
\centering
\epsfxsize=7.5 cm
\epsfbox{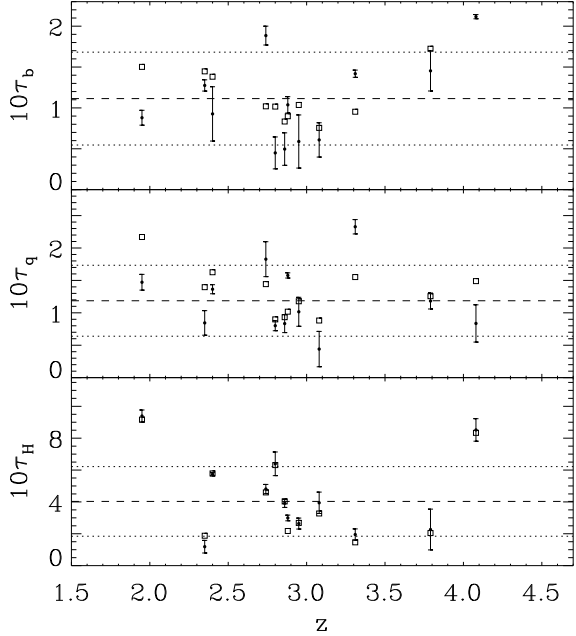}
\vspace{0.8cm}
\caption{ Redshift dependence of the smoothed amplitude, $(1+z)\tau$,
vs. $z$ for the sample of 12 QSO. Three definitions of DM density
are used: $\rho_{DM}\propto b^{3/2}$ (top panel), $\rho_{DM}\propto
(1+\eta^2)^{1/2}-1$ (middle panel), and $\rho_{DM}\propto N_{HI}$
(bottom panel). Open squares show the same amplitudes obtained for a
random redshift distribution of the same absorbers. The mean values
are plotted by the dashed lines and the dispersions are shown by
dotted lines.
}
\label{aut_f}
\end{figure}

For separate QSO the observed absorbers distribution along a line
of sight was averaged with the Gaussian window function and the
smoothed density was found as
$${\rho(x)\over\langle\rho\rangle} = \frac{1}{W_n(x)}\sum_i {q_i\over
\langle q\rangle}W(x-y_i),
						\eqno(6.1)$$
$$W(y-x)=\frac{1}{2\pi r_s}\exp\left(-\frac{(y-x)^2}{2r_s^2}\right),
~~W_n=\int_0^D dy W(x-y),$$
where $y_i \& q_i$ are coordinates and DM column density of
absorber, and $W_n$ is a normalization factor taking into account 
the finite size  of observed spectra. The results obtained for
absorbers with $N_{HI}\geq 10^{13}cm^{-2}$ and averaged
over 15 $r_s$, 7.5$\Omega_m^{-1/2}h^{-1}$Mpc $\leq ~r_s~\leq
40\Omega_m^{-1/2}h^{-1}$Mpc, are plotted in Fig. 6 together with
the dispersions plotted as error bars. Averaging over 12 QSO
gives for the three used definitions of $q$:
$$\langle (1+z)\tau(z)\rangle\approx (0.11\pm 0.06)
(\Theta_m)^{1/4},$$
$$\langle (1+z)\tau(z)\rangle\approx (0.12\pm 0.05)
(\Theta_m)^{1/4},				\eqno(6.2)$$
$$\langle (1+z)\tau(z)\rangle\approx (0.40\pm 0.22)
(\Theta_m)^{1/4}.$$
where, as before, $\Theta_m=9\Omega_mh^2$. The large scatter of
points shows that the used statistics of absorbers is insufficient
and does not allow to obtain reasonable estimates of the amplitude.

These results show that first and second definitions of DM
column density lead to similar estimates of $\tau$, whereas the third
definition increases it about 4 times. 

It is interesting to compare these results with similar estimates
obtained for randomly distributed absorbers with the same $b$ and
$N_{HI}$. Results averaged over 100 random realizations are
plotted in Fig. 6. For the first and second definitions
of DM column density, $q$, relative to the mean value, does not
differ by more than a factor of 2 -- 3  and the use of random 
redshift decreases the
dispersion about 2 times.  For the third definition variations of
$q$ around the mean value are much larger and can exceed a factor of 
$\sim$ 100.  Such a large discrepancy is a result of 
small fraction of absorbers with largest $N_{HI}$. 

This estimates of $\tau_q\approx\tau_b$ differ by about  2 -- 3 
times in comparison with the 
values $\tau_b$ obtained in Sec. 4.1. This divergence can be
caused by various factors, the most important are probably the
small representativity and relatively small size of spectra used,
what increases the uncertainty of all estimates for larger smoothing
scales. 

These results, as well as estimates of $\tau_b$
presented in Sec. 4.1, demonstrate the limited abilities of
investigations of properties of absorbers with respect to the
description of DM distribution. Both theoretical analysis
and investigations of simulations (Governato et al. 1998;
Jenkins et al. 1998) show that at redshifts $z\sim$ 3 high
density filaments accumulate main fraction of galaxies and
are the most typical and conspicuous structure elements. But
such filaments as well as separate galaxies constitute
relatively small fraction of the observed absorbers because of
their relatively small size, and are represented mainly by the
lower density halo. This means that the observed absorbers
give us more information about large low density regions
of the universe, and estimates of general characteristics of
spatial matter distribution derived from such analysis should 
be essentially corrected.

Results obtained above illustrate the potential and limitations 
of this approach rather then give the actual estimates of density 
variations.

\section{Summary and Discussion.}

In this paper the observed parameters of Ly-$\alpha$ lines
are analyzed and interpreted in the framework of the simple
self-consistent theoretical model of the DM structure
evolution (DD99). It is shown that the observed evolution of
absorbers is sensitive to several random factors and probably
does not trace directly the evolution of DM component. Our
results show however that some essential observed
characteristics of absorbers can be reasonably described
even by the statistical model considered here. 

The main results of our analysis can be summarized as follows:
\begin{enumerate}
\item The basic observed properties of Ly-$\alpha$ absorbers
	are satisfactorily described by the discussed
	statistical model of DM confined structure elements.
\item The observed characteristics of Doppler parameter, $b$,
	could be linked to the column density of accompanied 
	DM structure elements. It allows us to explain the 
	observed distribution of Doppler parameter which is 
	consistent with the Gaussian distribution of initial 
	perturbations. The measured amplitude of perturbations, 
	$\tau_b$, as given by (4.4), is consistent with that is 
	expected for lower density cosmological models.
\item The existence of observed galaxy and simulated DM walls 
	as well as the theoretical arguments demonstrate that 
	the merging of structure elements is one of the most 
	important factors of structure evolution. The observed 
	characteristics of entropy, $F_S$~\&~ $\Sigma$, confirm 
	that such a merging can be also considered as probably 
	the main factor of absorbers evolution at redshifts $z\geq 2$. 
\item For the main fraction of absorbers, the weak correlation 
	of column density, $N_{HI}$, with Doppler parameter, $b$,
	can be reasonably explained by the influence of variations  
	of entropy of compressed gas. The smooth observed distribution 
	of stronger absorbers, with $N_{HI}\geq 10^{14}cm^{-2}$, 
	remains in question and must be investigated, first of all, 
	with larger simulations. 
\item The redshift distribution of absorbers does not repeat the
	spatial distribution of galaxies and characterizes mainly
	the matter distribution within larger lower density
	regions. This makes it difficult to reconstruct the 
	spatial DM distribution from the redshift distribution 
	of absorbers. 
\end{enumerate}

The main statistical characteristics of absorbers distribution
discussed above coincide with published estimates (see, e.g.,
Hu et al., 1995; Cristiani 1995, 1996;  Kim et al. 1997). They
are roughly consistent with the expected evolution of DM structure
and can be, in principle, used to test and to discriminate
different models of structure evolution. At redshifts $z\geq 3.5$ 
the Doppler parameter and the mean linear number density of stronger 
absorbers are more sensitive to the influence of the initial power 
spectrum of perturbations and to the basic parameters of cosmological 
model.

The characteristics of DM structure elements discussed in Sec.
2.2 and 4.1 were tested against simulated DM distribution at
z=0 (DD99, DMRT, DDMT). The precision reached was $\sim$ 10\%.
Further application of these methods to broader set of
simulations at high redshifts will help to estimate the unknown
numerical factors and to improve the proposed here description. 
The discussed model of structure evolution is limited and cannot 
yet describe, for example, the structure disruption. It should be 
improved at least in this respect.

The available database cannot yet provide a precision required
for the reliable discrimination of models of structure formation.
Now the observational data are concentrated mainly in the
narrow range of redshifts $2.2< z <3.5$ and therefore  
the quantitative description of absorbers evolution is difficult. The wider
range of observed redshifts is required to improve the description
and to obtain more detailed and reliable information about the
structure evolution and physical processes at high redshifts.

\subsection*{Acknowledgments}
This paper was supported in part by Denmark's
Grundforskningsfond through its support for an establishment of
Theoretical Astrophysics Center, grant INTAS-93-68 and by the Polish 
State Committee for Scientific Research grant Nr. 2-P03D-014-17.  
AGD and VIT also wish to acknowledge support
from the Center of Cosmo-Particle Physics, Moscow.
Furthermore, we wish to thank the anonymous
referee for many useful comments.

\medskip
\centerline{\bf Appendix A}
\medskip
\centerline{\bf COBE normalized power spectrum}
\medskip

A convenient universal normalization of the power spectrum can be
obtained using the anisotropy of Microwave Background Radiation
measured by COBE. The convenient parametrisation applied to open
universe with matter density $\Omega_m\leq$ 1 and for the
spatially flat universe with $\Omega_m+\Omega_\Lambda=1$ can be
taken from Bunn~\&~White (1996).

For the Harrison - Zel'dovich initial power spectrum with the
BBKS transfer function we have:
$$\sigma^2_\rho(z) = 3\tau^2(z){m_0\over m_{-2}^2},\eqno(A.1)$$
where a usual definition of moments of power
spectrum is used:
$$m_{n} = \int_0^\infty x^{3+n}T^2(x)dx .\eqno(A.2)$$
The function $\tau(z)=\tau_0 B(z)$ depends on the model of
the universe and can be expressed trough $\Omega_m~\&~h$ as
follows:
$$\Omega_m+\Omega_\Lambda=1$$
$$B(z)\approx \left[{1-\Omega_m+2.2\Omega_m(1+z)^3\over
1+1.2\Omega_m}\right]^{-1/3},$$
$$(1+z)\tau(z)\approx 2.73h^2\Omega_m^{1.21}
[0.45\Omega_m^{-1}+0.55]^{1/3},~z\gg 1,\eqno(A.3)$$
and for $\Omega_m=0.3, ~h=0.6$ we have
$$(1+z)\tau(z)\approx 0.29.$$
For the open universe with $\Omega_\Lambda=0,~~\Omega_m\leq 1$
$$B^{-1}(z)\approx 1+{2.5\Omega_m\over 1+1.5\Omega_m}z,$$
$$(1+z)\tau(z)\approx 1.1h^2\Omega_m^{0.65-0.19ln\Omega_m}
(1+1.5\Omega_m),~z\gg 1\eqno(A.4)$$
and, for example, for $\Omega_m=0.5, ~h=0.6$ we have
$$(1+z)\tau(z)\approx 0.4 .$$

More details can be found in DD99 and DDMT.

\medskip
\centerline{\bf Appendix B}
\medskip
\centerline{\bf Bulk heating and cooling of gas}
\medskip

The thermal evolution of gas is described by the well known 
equation
$${3\over 2T}{dT\over dt} = {1\over n_b}{dn_b\over dt}+
{\epsilon_\gamma\Gamma_\gamma n_{HI}-
n_b^2(\beta_{fb}+\beta_{ff})\over n_bk_BT} ,                 \eqno(B.1)$$
where $\epsilon_\gamma$
is the mean energy injected at a photoionization
and radiative cooling coefficients for free-bound and
free--free emissions can be taken as
$$\beta_{fb} \approx 3.7~10^{-25}(T/10^4K)^{1/2}erg~cm^3/s,$$
$$\beta_{ff}\approx 2\cdot 10^{-25}(T/10^4K)^{1/2}erg~cm^3/s.$$
If the ionization equilibrium is reached and
$$\alpha_{rec} n_b^2 = n_{HI}\Gamma_\gamma$$
then equation (B.1) can be rewritten more transparently as
$${3\over 2T n_b}{dT\over dt} = {1\over n_b^2}{dn_b\over
dt}+\beta_s (T_\gamma/T-1),                      \eqno(B.2)$$
$$\beta_s = {\beta_{fb}+ \beta_{ff}\over k_BT}\approx
4.2\cdot 10^{-13}(T/10^4K)^{-1/2}cm^3s^{-1}$$
and $T_{\gamma}= (5 - 10)\cdot10^4K$ for the
suitable spectrum of UV radiation (Black, 1981).
$T_{\gamma}$ depends on the spectrum of local radiation
and can vary from point to point. This heating becomes
negligible for the cold high density clouds when the
radiative cooling due to the line emission is essential.
For the main population of observed absorbers
$$b_{obs}\sim{\rm25 - 30 km/s},\quad T_{obs}\approx (4-6)
\cdot 10^4K\leq T_\gamma,$$
and the influence of bulk heating could be
essential.

Eq. (B.2) can be suitably rewritten as follows:
$${d\over dt}{T^{3/2}\over n_b} = T^{3/2}\beta_s (T_\gamma/T-1),
			                      \eqno(B.3)$$
that allows us to integrate it directly. Using the definition
of entropy function, $F_S$, as given by (2.20) we obtain:
$$F_S^{3/2}(z) = F_S^{3/2}(z_f)+\alpha_s\int_z^{z_f}dx
{H_0\over H(x)}{T_\gamma(x)-T(x)\over (1+x)T_{bg}},$$
$$\alpha_s = 1.5h^{-1}\zeta^3\Theta_{bar}.		\eqno(B.4)$$
where $z_f$ is the redshift of pancake formation.
This result is obtained under condition of ionization
equilibrium. In the simplest case of fixed temperatures,
$T(z) = {\rm const}\leq T_\gamma(z)= {\rm const}$, we have
$$F_S^{3/2}(z)=F_S^{3/2}(z_f)+\alpha_s~H_0[t(z)-t(z_f)]
{T_\gamma - T\over  T_{bg} },				\eqno(B.5)$$
that means the slow growth of entropy and drop of the
density of absorbers and $N_{HI}$.

The influence of the bulk heating can be enhanced by the
pancake disruption due to the clustering of both DM and
baryonic components. This process is usually accompanied
by the adiabatic reduction of temperature in expanded
regions that accelerates the bulk heating. In this case, however,
the parameters $b$ and $N_{HI}$ are strongly correlated. 
This means that the possible contribution of such 
heating is restricted at least for richer absorbers.

\end{document}